\documentclass[12pt]{article} 

\usepackage{amsfonts}
 \usepackage{amssymb}
 \usepackage{amsmath}

\def\pmx{\begin{pmatrix}}
\def\emx{\end{pmatrix}}
\def\bsq{\begin{subequations}}
\def\esq{\end{subequations}}
\def\be{\begin{eqnarray}}
\def\ee{\end{eqnarray}}
\def\bee{\begin{eqnarray*}}
\def\eee{\end{eqnarray*}}
\def\bal{\begin{align}}
\def\eal{\end{align}}

\def\bra{\langle}
\def\ket{\rangle}
\def\dg{\dagger}
\def\kb{ \ket \bra }

 \def\tr{\hbox{Tr} \,}
 \def\trp{\hbox{Tr} }

\def\nn{\nonumber}
\def\ot{\otimes}

\def\hil{{\mathcal H}}

\def\ls{{\rm LS}}

\newcommand{\proj}[1]{ | #1 \kb  #1|}

\title{Comment on ``Stronger subadditivity of
entropy''   by  \\ Lieb and Seiringer  {\em Phys. Rev. A}  {\bf 71}, 062329 (2005) }

    \author{Mary Beth Ruskai \thanks{Partially supported  by
 the National Security Agency (NSA) and
 Advanced Research and Development Activity (ARDA) under
Army Research Office (ARO) contract number 
     DAAD19-02-1-0065, and by the National Science
        Foundation under Grant  DMS-0314228.}
      \\ {\small Department of Mathematics,
Tufts University,
     Medford, MA 02155 USA} \\ 
    {\small  Marybeth.Ruskai@tufts.edu}}

\begin{document}

\maketitle

  \begin{abstract}
  We show how recent results of Lieb and Seiringer can be obtained
  from repeated use of the monotonicity of relative entropy under partial traces,
  and explain how to use their approach to obtain tighter bounds in many 
  situations.
      \end{abstract}

In  \cite{LS}, Lieb and Seiringer (LS) proved an inequality which
they view as stronger than the well-known strong subadditivity (SSA) of
quantum entropy in a form equivalent to the contraction of  relative
entropy under partial traces.   At first glance, this may seem inconsistent 
with recent work of Ibinison, Linden and Winter \cite{ILW}, who prove
that this contraction is the {\em only} inequality satisfied by relative entropy.
There is no real contradiction because, as  LS
acknowledge in \cite{LS},  their results can be derived {\em from} SSA 
in the form of the  monotonicity  of 
relative entropy under completely positive  trace preserving (CPT) 
maps \cite{Lind75,OP,Uhl2}.
Nevertheless, it seems worth restating the results in \cite{LS} in a way 
that makes clearer the connection with montonicity of relative entropy.

We need some notation.   A density matrix
is a positive semi-definite matrix $\rho $ satisfying $\tr \rho = 1$.
 The entropy of a density matrix $\rho$ is given by $S(\rho) = - \tr \rho \log \rho$,
 and the relative entropy of a pair of density matrices $\rho, \gamma$ with
 $\ker(\gamma) \subset \ker(\rho)$ is given by
 \be
     H(\rho, \gamma) = \tr \rho  \, \big( \log \rho - \log \gamma \big).
 \ee
SSA can be written as an inequality for the conditional entropy
in the form
\be   \label{ssa}
   S(\rho_{BC}) - S(\rho_B)  \geq S(\rho_{ABC}) - S(\rho_{AB}) 
\ee
where $\rho_{ABC}$ is a density matrix on the
tensor product space $\hil_A \ot \hil_B  \ot \hil_C$, and the
reduced density matrices are given by
$\rho_{AB} = \trp_C \, \rho_{ABC}$, $\rho_B = \trp_A \, \rho_{AB} = \trp_{AC} \,  \rho_{ABC}$, etc.
Since the  conditional entropy  $S(\rho_{AC}) - S(\rho_A)$
  satisifies
\be
    S(\rho_{AC}) - S(\rho_A) = - H(\rho_{AC}, \rho_A \ot  \tfrac{1}{d_C} I_C) + \log d_C
\ee
with $\log d_C$ the dimension of $\hil_C$, \eqref{ssa} can be rewritten as
\be   \label{mpt}
  H(\rho_{BC}, \rho_B \ot  \tfrac{1}{d_C} I_C)   \leq  H(\rho_{ABC}, \rho_{AB} \ot \tfrac{1}{d_C} I_C) .
\ee
The term $ \tfrac{1}{d_C} I_C$ plays no role, except to ensure that the
second argument of $H( \cdot, \cdot)$ is a density matrix.

Given a set of operators  $\{ K_m \}$ satisfying $\sum_m K_m^{\dg} K_m = I$, 
Lieb and Seiringer \cite{LS} obtained an entropy inequality which
can be stated in terms of a CPT map we call $\Lambda_{\ls}$
and write as
\be    \label{LSdef}
  \Lambda_{\ls}(\rho) = \sum_m K_m \rho K_m^{\dg} \ot  \proj{m} 
\ee
Thus, $  \Lambda_{\ls}(\rho)$ is a block diagonal matrix with 
diagonal blocks  $K_m \rho K_m^{\dg}$.     The condition $\sum_m K_m^{\dg} K_m = I$
implies that the map 
\be  \label{kraus}
  \Phi(\rho) = \sum_m K_m \rho K_m^{\dg} .
\ee
 is  also a  CPT map.
By a slight modification\footnote{For details, see the Appendix of \cite{KMNR} and note that the representation above is equivalent to the familiar one \cite{KW} using
a unitary $U_{ADE}$ and a pure state $\proj{\phi_{DE}}$ such that 
$\sigma_{ADE}(\rho) = \linebreak U_{ADE} \, \big( \rho \ot \proj{\phi_{DE}} \big) \, U_{ADE}^\dag $.} of the standard Lindblad-Stinespring ancilla representation
 \cite{KMNR,KW,Lind75,Paul,Stine} of a CPT map, we   can represent both  
$\Phi$ and $\Lambda_\ls$ as partial traces
on the same extended space $\hil_A \ot \hil_D \ot \hil_E$  
(with $\hil_A$  the original Hilbert space) by defining
\be  \label{LSanc}
    \sigma_{ADE}(\rho) = \sum_{mn}  K_m \rho K_n^{\dg}  \ot |m\kb n|  \ot |m\kb n| = V \rho V^{\dg}
\ee
where $V $ is a block column vector with elements $K_m \ot | m \ket \ot | m \ket$.
It is easy to check that $V^{\dg} V = I$ so that $V$ is a partial isometry.
Then 
\be  \label{LSred}
 \Lambda_\ls(\rho)  =  \trp_{E} \, \sigma_{ADE}(\rho) 
\ee and 
\be
 \Phi(\rho)  = \trp_{DE} \, \sigma_{ADE}(\rho) =   \trp_D \, \Lambda_\ls(\rho).
\ee
We similarly define $ \tau_{ADE}(\gamma) = V \gamma V^{\dag}$. 
Since $V^{\dg} V = I$, one finds $S(\sigma_{ADE}) = S(\rho)$  and  
$H[ \sigma_{ADE}, \tau_{ADE}]  = H(\rho,\gamma)$, where we now
suppress argument in $\sigma_{ADE}(\rho)$ etc.   Therefore,
\be   \label{LSpre}
    H[\sigma_A, \tau_A]  ~\leq ~H[ \sigma_{AD}, \tau_{AD}] 
    ~    \leq   ~ H[ \sigma_{ADE}, \tau_{ADE}] 
\ee
is equivalent to
\be   \label{LSgen}
    H[\Phi(\rho), \Phi(\gamma)]  ~\leq ~H[ \Lambda_\ls(\rho),  \Lambda_\ls(\gamma)] 
    ~    \leq   ~ H(\rho,\gamma)
\ee
for any pair of density matrices $\rho, \gamma$.   

When
  $\rho = \rho_{ABC}$ and $ \gamma= \rho_{AB}  \ot \tfrac{1}{d_C} I_C$  the
  inequality   \eqref{LSgen}  can be written as
\bsq    \label{LSnew}  \be   \label{LSa}
    S(\rho_{ABC}) - S(\rho_{AB})  & \leq  &
       S\big[( \Lambda_\ls \ot I_C)(\rho_{ABC})\big] - S\big[ \Lambda_\ls(\rho_{AB})\big] \\
        & \leq  & S\big[(\Phi_{AB} \ot I_C)(\rho_{ABC})\big] - S\big[(\Phi_{AB})(\rho_{AB})\big] . 
       \label {LSb}
  \ee  \esq
  The first inequality \eqref{LSa} is the main theorem in \cite{LS}.  

When $K_m$ acts nontrivially only on $\hil_B$, LS obtained the following
inequality, which is (9) in \cite{LS}.
\be    \label{LS9}
 S(\rho_{ABC})  - S(\rho_{AB})      & \leq &
     S\big[( \Lambda_\ls  \ot I_C)(\rho_{ABC})\big] - S\big[ \Lambda_\ls(\rho_{AB})\big]  \nn  \\
  &   \leq  & S(\rho_{AC}) - S(\rho_{A})  
   \ee
   and the claim of ``stronger'' subadditivity rests on \eqref{LS9}.
   In this situation,  \eqref{LSred} becomes
$ \Lambda_\ls(\rho_{ABC})  =  \trp_E \, \sigma_{ABCE}$ with 
$\sigma_{ABCE} = U_{BE} \, ( \rho_{ABC} \ot \proj{\phi_{E}} ) \, U_{BE}^{\dag} $.  Then,
as before,
\begin{align}
S(\rho_{ABC})  - S(\rho_{AB})  &  ~ = ~   \log  d_C  
      -  \, H(\sigma_{ABCE}, \sigma_{ABE} \ot \tfrac{1}{d}  I_C) \\
\intertext{and, since $ \sigma_{AC} =\trp_{BE} \, \sigma_{ABCE}   = 
  \trp_{BE} \, \rho_{ABC} \ot \proj{\phi_{E}}   = \rho_{AC}$,}
S(\rho_{AC}) - S(\rho_{A})  & ~ = ~   \log  d_C
   -  \, H(\sigma_{AC}, \sigma_{A} \ot \tfrac{1}{d}  I_C).
\end{align}
Thus,   \eqref{LS9} is equivalent to
\be  \label{fin}
   H(\sigma_{ABCE}, \sigma_{ABE} )    \geq    H(\sigma_{ABC}, \sigma_{AB} ) 
        \geq H(\sigma_{AC}, \sigma_{A} ) 
\ee
where we have suppressed $\ot \tfrac{1}{d}  I_C$ in the second argument.    In view
of \eqref{fin},  the inequality \eqref{LS9} seems best  viewed as an
application of successive uses of SSA.

It is worth noting that  \eqref{LS9} holds if $\Lambda_\ls$ is replaced by
 {\em any} CPT map  for $\hil_B$; it need not have the
special form \eqref{LSdef}.   Indeed,  when $ \Lambda_\ls$ and $\Phi_B$
are related as in \eqref{LSdef} and \eqref{kraus}, one finds
that  \eqref{LSb} implies
\be    \label{LS9imp}
 S(\rho_{ABC})  - S(\rho_{AB})      & \leq &
     S\big[( \Lambda_\ls  \ot I_C)(\rho_{ABC})\big] - S\big[ \Lambda_\ls(\rho_{AB})\big]   \nn  \\
  &   \leq  & 
   S\big[( I_A  \ot \Phi_B  \ot I_C)(\rho_{ABC})\big] - S\big[ I_A \ot  \Phi_B (\rho_{AB})\big]  \nn \\
      &   \leq  &   S(\rho_{AC}) - S(\rho_{A}) .
   \ee
 Whether  $ \Lambda_\ls$ or $\Phi_B$ yields a ``stronger'' bound depends on
 whether one is trying to find a lower bound for $S(\rho_{ABC})  - S(\rho_{AB}) $
 or an upper  bound  for $S(\rho_{AC}) - S(\rho_{A}) $.

This suggests another way in which the  results in \cite{LS} might be
used to tighten bounds in some situations.  Given any CPT map $\Phi$,
one can use its representation \eqref{kraus}
to construct another CPT map  $\Lambda_\ls$ as in \eqref{LSdef}.
These maps satisfy the  inequalities \eqref{LSgen} and \eqref{LSnew}. 
Although the second  inequality in each pair need
not  be strict, one expects that to be the generic situation when $\Phi$ does not
already have the form $\Lambda_\ls$. Since the representation \eqref{kraus}
is not unique, one can find a family of such bounds.  


In \cite{LbBull}, Lieb considered several natural   ways of extending
SSA to more than three parties and showed each was either an easy
consequence of SSA or false.   In \cite{Pipp}, Pippenger gave a formal criterion
for deciding whether or not an entropy inequality is ``new'', and  independent of SSA,
in terms of a convex cone of entropy vectors.    Subsequently, 
Linden and Winter \cite{LW} found a new entropy inequality in the case of four parties and
evidence \cite{ILW2} for another.    However, the results in \cite{ILW}  imply that none of 
these can give a new inequality for the relative entropy.   Thus, any  strengthening
of SSA in the ``sandwich'' sense of LS \cite{LS} must be reducible to the form \eqref{fin}.


  \end{document}